\documentclass[aip,jcp,reprint,amsmath,amsfonts,amssymb]{revtex4-1}

\usepackage{graphicx}
\usepackage{comment}

\newcommand{\half}{\frac{1}{2}}

\begin{document}


\title{On the role of interfacial hydrogen bonds in ``on-water'' catalysis} 



\author{Kristof Karhan}
\affiliation{Department of Chemistry, University of Paderborn, Warburger Str. 100, D-33098 Paderborn, Germany}
\affiliation{Institute of Physical Chemistry, Johannes Gutenberg University Mainz, Staudinger Weg 7, D-55128 Mainz, Germany}
\author{Rustam Z. Khaliullin}
\affiliation{Institute of Physical Chemistry, Johannes Gutenberg University Mainz, Staudinger Weg 7, D-55128 Mainz, Germany}
\author{Thomas D. K\"{u}hne}
\email{tdkuehne@mail.upb.de}
\affiliation{Department of Chemistry, University of Paderborn, Warburger Str. 100, D-33098 Paderborn, Germany}
\affiliation{Institute of Physical Chemistry, Johannes Gutenberg University Mainz, Staudinger Weg 7, D-55128 Mainz, Germany}
\affiliation{Center for Computational Sciences, Johannes Gutenberg University Mainz, D-55128 Mainz, Germany}


\date{\today}

\begin{abstract}
Numerous experiments have demonstrated that many classes of organic reactions exhibit increased reaction rates when performed in heterogeneous water emulsions. Despite enormous practical importance of the observed ``on-water'' catalytic effect and several mechanistic studies, 
its microscopic origins remains unclear. In this work, 
the second generation Car-Parrinello molecular dynamics method 
is extended to self-consistent charge density-functional based tight-binding in order to 
study ``on-water'' catalysis of the Diels-Alder reaction between dimethyl azodicarboxylate and quadricyclane. We find that the stabilization of the transition state by dangling hydrogen bonds exposed at the aqueous interfaces plays a significantly smaller role in ``on-water'' catalysis than has been suggested previously.
\end{abstract}


\maketitle 




\section{Introduction} \label{sec:Introduction}

Liquid water as a solvent offers great economic and environmental benefits to chemical industry because it is inexpensive, nontoxic and nonflammable at the same time \cite{GreenChemistry}. 
However, up until recently, organic synthesis has mainly neglected water for that purpose, since most organic molecules are nonpolar and therefore 
highly insoluble in water. 
Thus, most organic reactions have been conducted in nonpolar or polar organic solvents \cite{OrganicSolvent}. 
Nevertheless, this situation has been changing since the early work of Breslow, in which it has been noticed that the presence of water enhances Diels-Alder reactions of two nonpolar hydrophobic molecules \cite{JACS.102.7816, ACR.24.159, b:inwater1}. 
Since then it has been demonstrated that many classes of organic reactions are accelerated under so-called ``in-water'' conditions
 \cite{ChemRev.93.2023, Synthesis, ChemCommun.2001.1701, Tetra.60.3205, JACS.126.444, ChemRev.105.3095, CEJ.12.1312, ChemRev.109.725, ChemRev.110.6302}. The observed catalytic effect has been explained by the H-bond induced stabilization of the transition state \cite{JACS.113.7430, JACS.114.5440, JOC.59.803, JCS.90.1727, JSCPT2.1997.653, JCPB.106.8078, JOC.67.510, ChemRev.110.6302, JACS.126.11923} and the positive influence of hydrophobic forces, which pushes the reactant molecules towards each other \cite{JOC.59.5372, PAC.67.823, JOC.61.5492, JOC.63.8989, PAC.72.1365, OBC.1.2809, JOC.73.8723}, the high cohesive density of liquid water \cite{CEJ.12.1312, Desimoni1990}, as well as catalysis through Lewis and Br{\o}nstedt acids \cite{JACS.95.4094, JACS.112.4127, JACS.118.7702, EJOC.2001.2001, JPOC.17.180,CEJ.16.8972}. 
 
The use of water as a solvent for organic synthesis was further popularized by Sharpless and co-workers, who demonstrated that numerous important uni- and bimolecular organic processes, such as cycloadditions, ene reactions, Claisen rearrangements and nucleophilic substitutions, exhibit greatly increased reaction rates, enhanced selectivity and improved yields when performed in vigorously stirred aqueous emulsions \cite{ACIE.44.3275}. The observed effect is now widely described as ``on-water" catalysis in order to emphasize that an oil-water interface is essential for the enhanced reactivity \cite{Nature.435.746}. 

Designing synthetic processes that efficiently exploit the catalytic effect of ``on-water" reactions requires a thorough understanding of its physical nature. However, the origin of the increased efficiency remains unclear. This is mainly due to the fact that probing molecular-scale dynamic processes directly by experiment at interfaces is a formidable challenge, even with the most advanced surface-sensitive techniques \cite{ChemRev.102.2693, a:revwater, a:revsaykally, AccChemRes.42.1332}, which is why computer simulations are an indispensable tool to elucidate the underlying mechanism of ``on-water'' catalysis. The structure of interfacial water, however, have been studied extensively, both by experiment \cite{PhysRevLett.70.2313, Science.264.5160, Science.292.5518, JPCM.20.205105, JCP.135.021101, JACS.133.10360, JACS.133.16875, Nature.474.192, PCCP.16.7377} and theory \cite{JCP.80.4448, JCP.106.8149, CP.258.371, Science.303.658, JPCB.110.3738, JCTC.3.2002, JPCB.113.4125, JPCB.113.11672, JPCL.2.105, JCP.135.044701, JCP.135.124712, JPCL.4.83, JPCL.4.3245}. All of these studies consistently observe a sizable fraction of dangling (or free) OH bonds at the surface of liquid water, which protrudes out of the water phase. This immediately points to an increased reactivity of the water surface that may facilitate heterogeneous catalysis \cite{ChemRev.96.1445, Nature.303.5658}. 

\begin{figure}
    \includegraphics[width=0.4\textwidth]{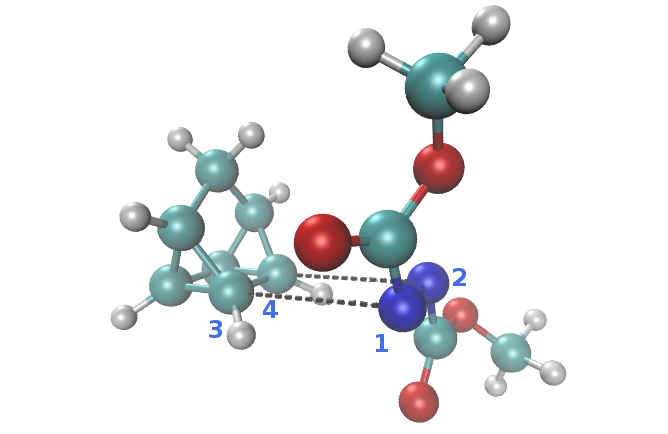}
	\caption{Schematic of the Diels-Alder reaction between Q (left) and DMAD (right). The dotted lines represent our reaction coordinate. The C, N, oxygen and H atoms are denoted by green, blue, red and white spheres, respectively.} \label{fig:DA-Reaction}    
\end{figure}

Inspired by this observation, Jung and Marcus 
conducted a theoretical investigation on the effect of interfacial water on the Diels-Alder reaction between quadricyclane (Q) and dimetyl azodicarboxylate (DMAD) (see Fig.~\ref{fig:DA-Reaction}) -- the most remarkable example of the ``on-water'' phenomenon \cite{JACS.129.5492}. 
They attributed the origin of ``on-water'' catalysis to the stabilization of the transition state through the formation of H-bonds between DMAD molecules and the free OH-groups that are exposed at the aqueous interface and protrude into the organic phase. Hence, contrary to the homogenous reaction in water solution, no H-bonds between the water molecules must be broken to yield free OH-groups, which can then act as a catalyst \cite{JACS.129.5492}. However, it has been argued that their model of the reaction environment, consisting only of the DMAD+Q complex plus three water molecules to mimic the surface, is an oversimplified representation of the actual water-organic interface \cite{JPCM.22.284117, JACS.132.3097}. Simulations in the condensed-phase or at finite temperature were not performed. As a consequence, subtle but important phenomena, such as assembly due to hydrophobic forces, as well as the dynamical nature of solvent-solute interactions at finite temperature were neglected. For that purpose, in a subsequent finite temperature study of Jorgensen and coworkers, the water surface was explicitly treated, though only at the QM/MM level of theory, where polarization effects of the solvent water molecules were not taken into account \cite{JACS.132.3097}. Moreover, the authors calculated the free energies for the Diels-Alder reaction of an isolated cyclopentadiene with three different dienophiles (1,4-naphthoquinone, methyl vinyl ketone and acrylonitrile) in bulk water, as well as on the water surface instead of the more realistic interface between the organic and the water phase. This comparative study suggests that the energy barriers of the homogeneous reactions are lower than those of the heterogenous ``on water'' reactions \cite{JACS.132.3097}, in contrast with the findings of Jung and Marcus \cite{JACS.129.5492, JPCM.22.284117}. 

The objective of the present paper is to 
further investigate the origin of ``on-water'' catalysis by means of finite temperature molecular dynamics simulations. For that purpose we explicitly consider the extended water-organic interface, which represents a more realistic model system of the oil emulsion found in experiment. 
The quantum nature of the electrons is approximately treated by the self-consistent charge density-functional based tight-binding method (SCC-DFTB), which is capable of qualitatively reproducing electronic polarization effects of solute and solvent molecules, as well as the electron delocalization between them \cite{PRB.58.7260, JPCM.14.3015}. In order to make such large scale simulations feasible, we have extended the second generation Car-Parrinello molecular dynamics (CPMD) approach of K\"uhne et al. \cite{PRL.98.066401, JCTC.5.235} to the SCC-DFTB method, as detailed in the Appendix \ref{sec:ESProp}.

\section{Computational details} \label{sec:Methods}

All simulations were performed using the CP2K suite of programs \cite{CPC.167.103}, where the present SCC-DFTB based second generation CPMD scheme has been implemented. As outlined in Appendix \ref{sec:ESProp}, this method unifies the efficiency of the CPMD approach with the large integration time steps of Born-Oppenheimer molecular dynamics (BOMD) \cite{PhysRevLett.55.2471, CP2Greview}. At variance to the recently proposed SCC-DFTB based BOMD scheme of Karplus and coworkers \cite{JCP.135.044122}, not even a single diagonalization step, but just an electronic force calculation is required, similar to the SCC-DFTB based CPMD technique of Seifert and coworkers \cite{JCP.126.124103}. However, contrary to the latter, integration time steps up to the ionic resonance limit can be taken that are typically one order of magnitude larger. Beside the original SCC-DFTB parameters \cite{PRB.58.7260}, modified parametrizations for the N-H and water interactions were employed \cite{JACS.126.14668, DFTBparam}. Our model of the homogeneous ``on water'' reactions consisted of a single DMAD+Q complex on a 360 molecule water slab that we denote as the water-vacuum reaction, whereas the water-organic system comprised of 19 DMAD+D placed on top of a 286 molecule water slab. For the sake of comparison we have also conducted a homogenous ``on water'' reaction of a single DMAD+Q complex solvated in 197 water molecules and the corresponding gas phase reaction of just one isolated DMAD+Q molecule. The corresponding simulations were conducted in the canonical NVT ensemble at $T=300$~K, where the volume of the orthorhombic or cubic simulation box was chosen to yield a density of 0.98~g/cm$^3$ and 1.02~g/cm$^3$ for the water and organic phases, respectively. Three dimensional periodic boundary conditions were employed throughout with an additional vacuum portion of $\sim$10~nm for each isolated direction. The Langevin dynamics was integrated using the algorithm of Bussi and Parrinello \cite{PRE.75.056707} together with a friction coefficient of $\gamma_L = 1/25$~fs$^{-1}$ and a discretized time step of $\Delta t=0.5$~fs. The simultaneous propagation of the electronic degrees of freedom occurred with $K=5$, which corresponds to a time reversibility of $\mathcal{O}(\Delta t^{8})$. Since this was disordered system at finite temperature that also exhibits a large band gap, the Brillouin zone is sampled at the $\Gamma$-point only.

The distances $d_{13}$ and $d_{24}$ were selected to be the reaction coordinate, as shown in Fig.~\ref{fig:DA-Reaction}. For the sake of simplicity, both distances were assumed to be identical, i.e. $d_{13}=d_{24}$. The free energy profile along the reaction coordinate was computed by constrained MD as an ensemble average of the Lagrangian multipliers using the \textsc{rattle} algorithm \cite{JCP.109.7737, JCoP.52.24}. 
To that extend, the reaction coordinates was discretized and the corresponding potential of mean force (PMF) evaluated for 44 distances from 1.52~\r{A} (product) to 5~\r{A} (isolated reactants). 
First, the gas phase reaction had been studied by equilibrating each of the 44 replicas of the system for 100~ps, before accumulating statistics for additional 200~ps. All replicas were then solvated in liquid water and placed in close vicinity to the water surface for the water-vacuum, as well as water-organic simulations. 
For all the systems, each replica was then again equilibrated for 50~ps, before averaging the Lagrangian multipliers for yet another 100~ps to yield the corresponding PMFs. The H-bond network around the solvated molecules was analyzed using a geometry-based definition as obtained by the technique of Skinner and coworkers \cite{JCP.126.204107}. We have verified that this definition yield an average of 3.7 H-bonds per molecule in bulk water and 2.8 at the surface, which is in close agreement with experiment \cite{JPCL.2.105}. 


\section{Results and discussion} \label{sec:Results}

\begin{figure}
    \includegraphics[width=0.5\textwidth]{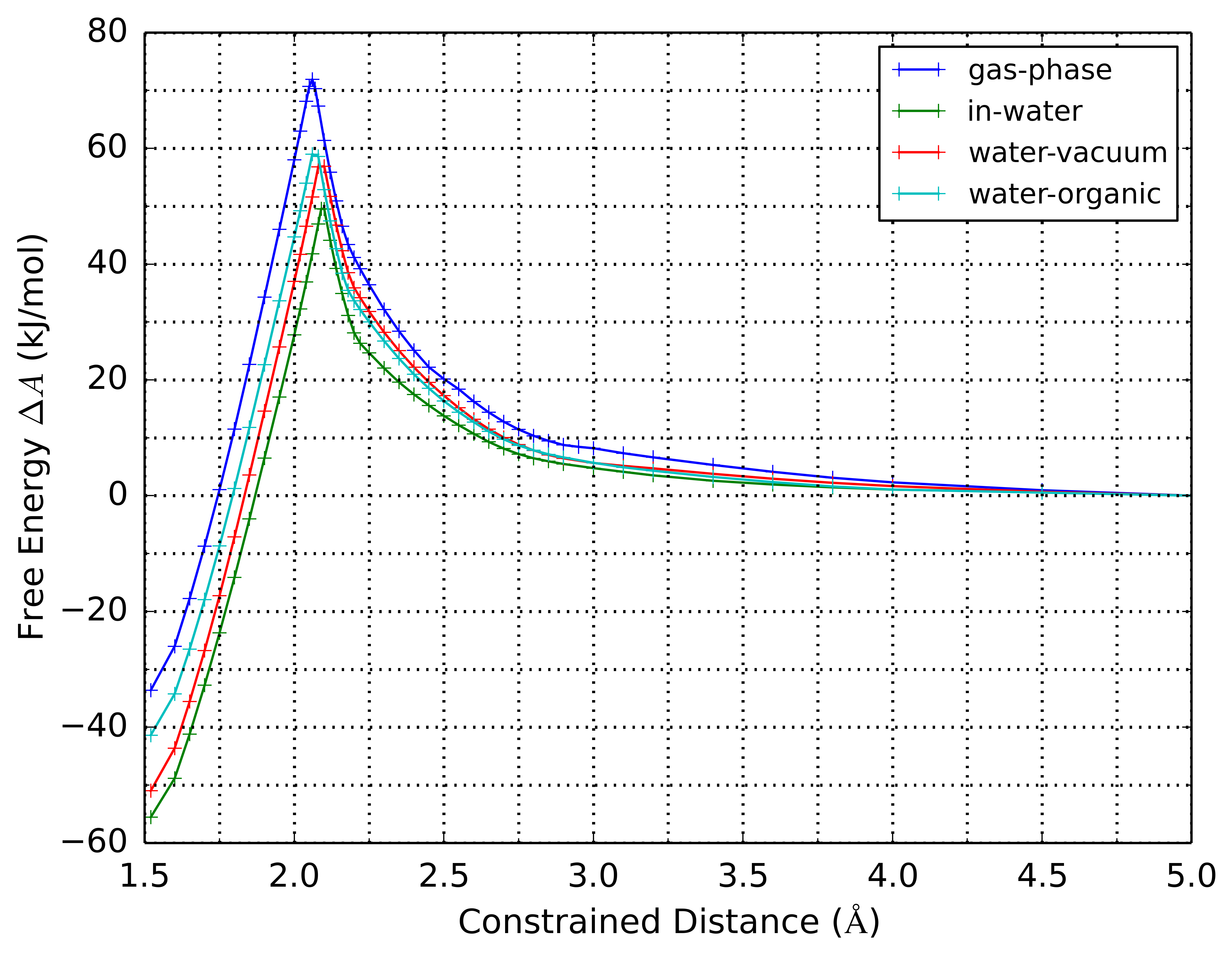}
    \caption{The free energies profiles for the Diels-Alder reaction between Q and DMAD in various environments. The specific free energy barrier heights $\Delta A^\ddagger$ are 72~kJ/mol for the gas-phase, 50~kJ/mol for the homogeneous ``in-water'', 57~kJ/mol and 59.18~kJ/mol for the heterogeneous water-vacuum, as well as water-organic ``on-water''reactions, respectively.}
    \label{fig:wv-wo-pmf}
\end{figure}

The free energy barriers for the various reaction environments are shown in Fig.~\ref{fig:wv-wo-pmf} and vary from 50 to 72~kJ/mol, which is in qualitative agreement with previous results for the Diels-Alder reaction calculated by others \cite{JACS.129.5492, JPCM.22.284117, JACS.132.3097}.
In particular, the quantitative agreement with the potential energy barriers calculated at the density functional level of theory (56~kJ/mol for the homogenous ``in-water'' reaction and 77~kJ/mol for the reaction in the gas-phase \cite{JOC.73.8791}) demonstrates the validity of the present approach. Our calculations show that the free energy barriers of the water-vacuum system is between the gas-phase and homogeneous ``in water'' reactions. In fact, the barrier of the former is $\sim$10~kJ/mol higher than that of the latter, which is consistent with the results of Jorgensen and coworkers \cite{JACS.132.3097}, but at variance with the findings of Jung and Marcus \cite{JACS.129.5492, JPCM.22.284117}. 
The free energy barrier of the water-organic reaction, however, is even higher by $\sim$2~kJ/mol than the one of the water-vacuum system. 

\begin{figure}
    \includegraphics[width=0.49\textwidth]{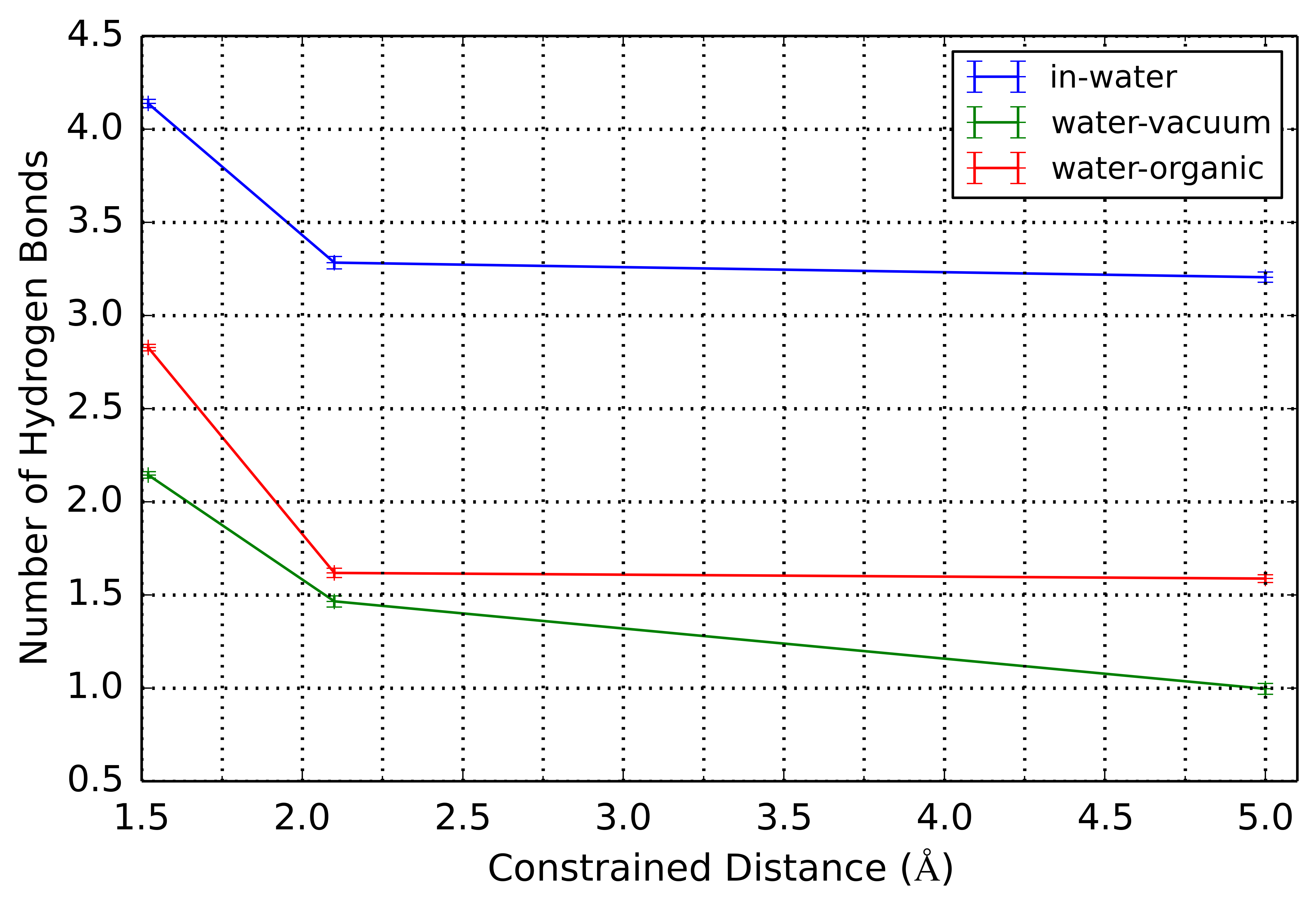}
    \caption{The average number of H-bonds between the polar N and O atoms of the DMAD complex and the solvent water molecules along the reaction coordinate. The error bar refers to the standard error.} 
    \label{fig:HBonds}
\end{figure}

To elucidate the impact of the various solvation environments on the reaction barriers, the average number of H-bonds between the negatively charged N and O atoms in the DMAD complex and the solvent water molecules has been calculated along the reaction coordinate, as shown in Fig.~\ref{fig:HBonds}.
In agreement with previous simulations of Jorgensen and coworkers, we also observe a significant loss of H-bonds for the reactants and transition states when changing from bulk water to the water-vacuum interface \cite{JACS.132.3097}. Whereas for the water-vacuum case the mean number of H-bonds is reduced by more than 2, the water-vacuum reaction is experiencing a slightly smaller reduction of $\sim$1.5~H-bonds. Since the deficit of H-bonds mitigates the hydrophobicity-induced \textit{internal pressure} effect \cite{JACS.102.7816, ACR.24.159}, the reaction barriers of both ``on-water'' reactions is therefore higher than ``in-water''. However, in their work, Jung and Marcus suggested that this is compensated by an increase of H-bonds at the transition state relative to the reactants, which is made possible by the large number of free OH-bonds at the water surface and thus results in an additional stabilization of the transition state \cite{JACS.129.5492, JPCM.22.284117}. However, as shown in Fig.~\ref{fig:HBonds}, our simulations suggest that at the transition state of the ``in-water'' and  water-organic reactions, the number of H-bonds is only marginally increased. While the number of H-bonds increases by 0.08 for the homogenous reaction, the latter increases by a mere 0.03~H-bonds only. Only for the water-vacuum system, there is an increase of 0.46~H-bonds, which is why in this case the free energy barrier is somewhat lower than that of the water-organic reaction in spite of the smaller overall number of H-bonds. Nevertheless, this is a direct consequence of the fact that the transition structure is small enough to maximize the number of H-bonds by being more buried in the surface that shifts the water-vacuum free energy barrier below the one of the water-organic reaction.

\section{Conclusions}

Our simulations support the general guiding principle of Jung and Marcus that the dangling OH-bonds at the water-organic interface affects the energetics of the reaction between DMAD and Q \cite{JACS.129.5492}. However, they also imply that the stabilization of the transition state by H-bonding is less relevant than has been previously suggested: the mere presence of the reactants at the interface does not reduce the free energy barrier of the reaction compared to the homogeneous "in-water" conditions. In our simulations, we do not observe additional stabilization from the increased number of H-bonds formed by transition states at the interface. We conclude by noting that our results do not exclude the possibility that certain preferential orientations or the increased strength of hydrogen bonds formed by the transition state can lead to the faster reaction at the water-organic interface. 
A detailed study on the stabilization effect of H-bonds based on a novel energy decomposition analysis method \cite{NatC.4.1450, a:khal-review-pccp}, will be reported elsewhere.
\begin{acknowledgments}
The authors would like to thank Yousung Jung and Johan Albrecht Schmidt for many fruitful discussions. Financial support from the Graduate School of Excellence
MAINZ and the IDEE project of the Carl Zeiss Foundation is kindly acknowledged. R.Z.K. is gratefully financial support from the Swiss National Science Foundation, while T.D.K. appreciates the Gauss Center for Supercomputing (GCS) for providing computing time through the John von Neumann Institute for Computing (NIC) on the GCS share of the supercomputer JUQUEEN at the J\"ulich Supercomputing Centre (JSC).
\end{acknowledgments}


\appendix

\section{SCC-DFTB based second generation CPMD} \label{sec:ESProp}

In the SCC-DFTB method \cite{PRB.58.7260, JPCM.14.3015}, the total potential energy of a closed-shell system is denoted as
%
\begin{eqnarray}
    \label{eq:EDFTB}
        E&=&Tr\left[\mathbf{C}^T\mathbf{H}^0\mathbf{C}\right] + \half\sum_{I,J}V_{IJ}(R_{IJ}) \nonumber \\
        &+&\half\sum_{I,J}\gamma_{IJ} \Delta q_I \Delta q_J,
\end{eqnarray}
where $\mathbf{H}^0=\mathbf{H}[\rho_0]$ is the zeroth-order Hamiltonian matrix that is a functional of the density $\rho_0$, which corresponds to the superposition of neutral atomic charge densities. As is customary, the expansion coefficients $C_{\mu i}$ of the $N_{occ}$ lowest occupied atomic orbitals $\phi_{\mu}$ are arranged in a rectangular matrix $\mathbf{C}$ and $\psi_i=\sum_{\mu} C_{\mu i} \phi_{\mu}$ are the molecular orbitals. The elements of $\mathbf{H}^0$ are tabulated together with the overlap matrix elements $S_{\mu\nu}=\langle\phi_{\mu}|\phi_{\nu}\rangle$ with respect to the interatomic distance $R_{IJ}$. The repulsive pair potential $V^{rep}(R_{IJ})$ depends only on $R_{IJ}$ and is fitted to approximate the so-called ``double counting'' correction terms of density functional theory (DFT) plus the nuclear Coulomb interaction. The remaining term is to mimic the second-order contribution of the DFT energy functional, where $\gamma_{IJ}$ describes the effective on-site electron-electron interaction and is related to the chemical hardness $\eta_{I}$, or the Hubbard parameter $U_{I}$, i.e. $\gamma_{II} \approx 2\eta_{I} \approx U_{I}$. The charge fluctuations are estimated by Mulliken charges 
\begin{equation}
    \Delta q_I=\sum_{i}\sum_{\mu\in I}\sum_{\nu} C_{\mu i} C_{i \nu}^T S_{\mu\nu}, 
\end{equation}
where $Z_I$ is the nuclear charge. Due to the fact that $\Delta q_I$ depends on $\mathbf{C}$, the solution of the generalized eigenvalue problem $\sum_{\nu}{C_{\nu i}(H^0_{\mu \nu} - \varepsilon_{i} S_{\mu \nu}) = 0}$, the charges are adapted by a self-consistent field (SCF) procedure \cite{JCP.140.134109}. 

An analytic expression for the interatomic forces can be easily derived by differentiating Eq.~\ref{eq:EDFTB} with respect to the nuclear coordinates $\mathbf{R}_I$ i.e.
\begin{eqnarray}
  \mathbf{F}_{I} &=& - \sum_{i}\sum_{\mu \nu}{C_{\mu i} C_{i \nu}^T \left[ \frac{\partial H_{\mu \nu}^0}{\partial \mathbf{R}_I} - \left( \varepsilon_i - \frac{\partial H_{\mu \nu}^1}{\partial S_{\mu \nu}} \right) \frac{\partial S_{\mu \nu}}{\partial \mathbf{R}_I} \right]} \nonumber \\
  &-& \Delta q_{I} \sum_{J}{\frac{\partial \gamma_{IJ}}{\partial \mathbf{R}_I} \Delta q_{J}} - \frac{\partial E_{rep}}{\partial \mathbf{R}_I},
\end{eqnarray}
where $H_{\mu \nu}^1 = \half S_{\mu \nu} \sum_{K} (\gamma_{IK}+\gamma_{JK}) \Delta q_{K}$ is a second-order correction term in terms of a Mulliken charge dependent contribution to $H_{\mu \nu}^0$ and $E_{rep} = \half\sum_{I,J}V_{IJ}(R_{IJ})$.

In SCC-DFTB based BOMD, the potential energy of a system is fully minimized for each MD time step, which renders this approach computationally rather expensive. By contrast, the CPMD approach bypasses the expensive iterative minimization by considering the electronic degrees of freedom as classical time-dependent fields with a fictitious mass and a suitably designed electron-ion dynamics for their propagation \cite{PhysRevLett.55.2471, WIRES.2.604}. The fictitious mass has to be chosen small enough to separate the electronic and nuclear subsystems, as well as forces the electrons to adiabatically follow the nuclei. 
However, in CPMD the maximum permissible integration time step scales like the square root of the fictitious mass parameter, and has therefore to be significantly smaller than that of BOMD, thus limiting the attainable simulation timescales \cite{NumMa.78.359}.

The second-generation CPMD method combines the best of both schemes by retaining the large integration time steps of BOMD and, at the same time, preserving the efficiency of CPMD \cite{PRL.98.066401, a:khal-review-pccp, CP2Greview}. To that extent, the original fictitious Newtonian dynamics of CPMD is substituted by an improved coupled electron-ion dynamics that keeps the electrons very close to the instantaneous electronic ground state and does not require an additional fictitious mass parameter. The superior efficiency of this new approach, which, depending on the system, varies between one to two orders of magnitude, is due to the fact that not only the SCF cycle, but also the iterative wave function optimization is fully bypassed. In other words, not even a single diagonalization step is required, while, at the same time, remaining very close to the instantaneous electronic ground state and allowing for integration time steps as large as in standard BOMD.

\section{Coupled Electron-Ion Dynamics}

Since the dynamics of the contra-covariant density matrix $\mathbf{PS}$, where $\mathbf{P}=\mathbf{CC}^T$ is the one-particle density kernel,  is much smoother than the one of the more widely varying wave functions immediately suggests to propagate the $\mathbf{PS}$ instead of the $\mathbf{C}$ matrix as in CPMD. In second generation CPMD, this is achieved by adapting the always stable predictor-corrector (ASPC) integrator of Kolafa \cite{JCC.25.335,JCP.122.164105} to the electronic structure problem. Specifically, we write the predicted wave function at time $t_n$ in terms of the $K$ previous $\mathbf{PS}$ matrices as
%
\begin{eqnarray}
    \label{eq:Predictor}
        \mathbf{C}^p\left(t_n\right) &\cong& \sum_{m=1}^K \left(-1\right)^{m+1} m \frac{{2K\choose K-m}} {{2K-2 \choose K-1}}\underbrace{\mathbf{C}\left(t_{n-m}\right)\mathbf{C}^T \left(t_{n-m}\right)}_{\mathbf{P}\left(t_{n-m}\right)} \nonumber \\
        &\times& \mathbf{S}\left(t_{n-m}\right)\mathbf{C}\left(t_{n-1}\right),
\end{eqnarray}
where the propagated $\mathbf{PS}(t_n)$ matrix is utilized as an approximate projector onto the occupied subspace $\mathbf{C}\left(t_{n-1}\right)$. The modified predictor is followed by a corrector step to minimize the error and to further reduce the deviation from the instantaneous electronic ground state. Considering that this scheme was originally devised to deal with classical polarization, special attention must be paid to the fact that the holonomic orthonormality constraint of the orbitals is always satisfied during the evolution. Therefore, the $\mathbf{C}$ matrix must obey the condition $\mathbf{C}^T\mathbf{SC} = \mathbf{I}$, which is due to the fermionic nature of electrons that compels the wave function to be antisymmetric in order to comply with the Pauli exclusion principle. 
%
Because of that, the modified corrector
\begin{eqnarray}
    \label{eq:Corrector}
    \mathbf{C}\left(t_n\right)&=&\mathbf{C}^p(t_n) + \omega \left( \mathrm{MIN}\left[\mathbf{C}^p(t_n)\right] - \mathbf{C}^p(t_n) \right), \nonumber \\
    \text{where}~\omega&=&\frac{K}{2K-1}~\text{and}~K\geq 2,
\end{eqnarray}
involves the evaluation of just one preconditioned electronic gradient $\mathrm{MIN}[\mathbf{C}^p(t_n)]$, using an orthonormality conserving minimization technique, to calculate the electronic force. The present predictor-corrector scheme is very accurate and leads to an electron dynamics that is time reversible up to $\mathcal{O}(\Delta t^{2K-2})$, while $\omega$ is chosen to guarantee a stable relaxation toward the electronic ground state. The efficiency of this approach is such that the electronic ground state is very closely approached within just one electronic gradient calculation. We thus totally avoid the SCF cycle and any expensive diagonalization steps.

\section{Modified Langevin equation}

Despite the close proximity of the electronic degrees of freedom to the instantaneous ground state the nuclear dynamics is dissipative, most likely because the employed electron propagation scheme is not symplectic. However, it is possible to correct for the dissipation by devising a modified Langevin equation that in its general form reads as:
%
\begin{equation}
    M_I \Ddot{\mathbf{R}}_I =\mathbf{F}_I^{BO}-\gamma {M}_I\dot{\mathbf{R}}_I + \mathbf{\Xi}_I, 
    \label{GeneralizedLangevinEquation}
\end{equation}
where where $M_I$ are the nuclear masses, $\mathbf{R}_I$ the nuclear coordinates (the dot denotes time derivative), $\mathbf{F}^{BO}_I$ the exact but unknown Born-Oppenheimer forces, $\gamma$ a damping coefficient and $\mathbf{\Xi}_I$ an additive white noise, which must obey the fluctuation-dissipation theorem $\langle \mathbf{\Xi}_I(0) \, \mathbf{\Xi}_I(t) \rangle = 2\gamma M_I k_BT \delta(t)$ in order to correctly sample the canonical Boltzmann distribution. 

Given that the energy is exponentially decaying, which had been shown to be an excellent assumption \cite{PRL.98.066401, WIRES.2.604}, it is possible to rigorously correct for the dissipation, by modeling the nuclear forces arising from our dynamics as: 
\begin{equation}
    \mathbf{F}^{CP}_I=\mathbf{F}_I^{BO} - \gamma_D {M}_I \dot{\mathbf{R}}_I, 
    \label{LangevinModelForces}
\end{equation}
where $\mathbf{F}^{CP}_I$ are the nuclear forces from second generation CPMD, while $\gamma_D$ is an intrinsic, yet unknown damping coefficient to mimic the dissipation. 

By substituting Eq.~\ref{LangevinModelForces} into Eq.~\ref{GeneralizedLangevinEquation}, the desired modified Langevin-like equation is obtained:
\begin{equation}
\label{eq:Corr-Lan-Dyn}
{M}_I\ddot{\mathbf{R}}_I = F^{CP}_I + \mathbf{\Xi}_I.
\end{equation}
%
%
%
%

In other words, if one knew the unknown value of $\gamma_D$ it would nevertheless be possible to guarantee an exact canonical sampling of the Boltzmann distribution (in spite of the dissipation) by simply augmenting Eq.~\ref{eq:Corr-Lan-Dyn} with $\mathbf{\Xi}_I$ according to the fluctuationÐdissipation theorem. Fortunately, the intrinsic value of $\gamma_D$ does not need to be known \textit{a priori}, but can be determined in a preliminary run in such a way that eventually the equipartition theorem $\langle \half {M}_I \dot{\mathbf{R}}^2_I \rangle = \frac{3}{2} k_B T$ holds \cite{PRB.73.041105, PRL.98.066401}. Although this can be somewhat laborious, but once $\gamma_D$ is determined very long and accurate simulations can be routinely performed at a greatly reduced computational cost.

%
%

\bibliographystyle{apsrev4-1}

%

\end{document}